\documentclass[conference]{IEEEtran}
\IEEEoverridecommandlockouts
\usepackage{cite}
\usepackage{amsmath,amssymb,amsfonts}
\usepackage{algorithmic}
\usepackage{graphicx}
\usepackage{textcomp}
\usepackage{xcolor}
\def\BibTeX{{\rm B\kern-.05em{\sc i\kern-.025em b}\kern-.08em
    T\kern-.1667em\lower.7ex\hbox{E}\kern-.125emX}}

\usepackage[normalem]{ulem}

\usepackage{multirow}
\usepackage{epsfig}
\usepackage{epstopdf}
\usepackage{subfig}
\usepackage{caption}
\usepackage{url}
\usepackage{breakurl}

\begin{document}

\title{
XOS: An Application-Defined Operating System for Datacenter Computing}

\author{\IEEEauthorblockN{Chen Zheng\IEEEauthorrefmark{1}\IEEEauthorrefmark{2},
Lei Wang\IEEEauthorrefmark{1},
Sally A. McKee\IEEEauthorrefmark{3},
Lixin Zhang\IEEEauthorrefmark{1},
Hainan Ye\IEEEauthorrefmark{4} and
Jianfeng Zhan\IEEEauthorrefmark{1}\IEEEauthorrefmark{2}\thanks{The corresponding author is Jianfeng Zhan.}}
\IEEEauthorblockA{\IEEEauthorrefmark{1}State Key Laboratory of Computer Architecture, Institute of Computing Technology, Chinese Academy of Sciences}
\IEEEauthorblockA{\IEEEauthorrefmark{2}University of Chinese Academy of Sciences, China}
\IEEEauthorblockA{\IEEEauthorrefmark{3}Chalmers University of Technology}
\IEEEauthorblockA{\IEEEauthorrefmark{4}Beijing Academy of Frontier Sciences and Technology}
}

\maketitle

\begin{abstract}
Rapid growth of datacenter (DC) scale, urgency of cost control, increasing workload diversity, and huge software investment protection place unprecedented demands on the operating system (OS) efficiency, scalability,
performance isolation, and backward-compatibility. The traditional OSes are not built to work with
deep-hierarchy software stacks, large numbers of cores, tail latency guarantee, and
increasingly rich variety of applications seen in modern DCs,
and thus they struggle to meet the demands of such workloads.

This paper  presents XOS, an application-defined OS for modern DC servers. Our design moves resource
management out of the OS kernel, supports customizable kernel subsystems in user space,
and enables elastic partitioning of hardware resources. Specifically, XOS leverages modern
hardware support for virtualization to move resource management functionality out of
the conventional kernel and into user space, which lets applications achieve near
bare-metal performance. We implement XOS on top of
Linux to provide backward compatibility. XOS speeds up a set of DC workloads by up to 1.6$\times$
over our baseline Linux on a 24-core server, and outperforms the state-of-the-art Dune by up to 3.3$\times$ in terms of virtual memory management. In addition, XOS demonstrates good
scalability and strong performance isolation.

\end{abstract}


\section{Introduction}

\begin{figure}[tb]
{
\centering
    \subfloat[monolithic]{\includegraphics[width=0.44\columnwidth]{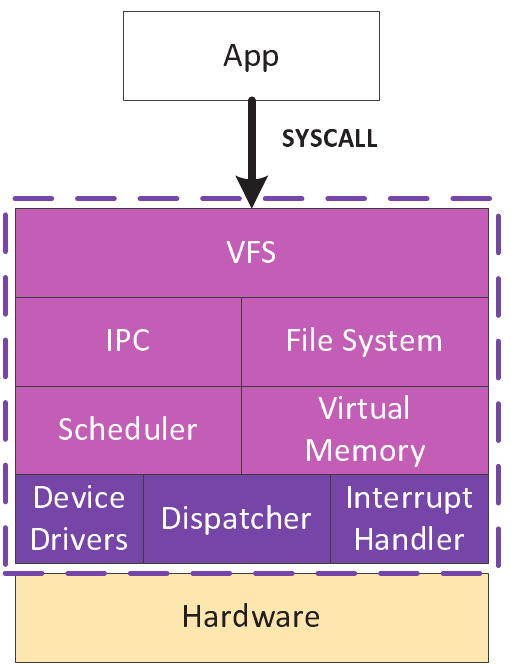}
	\label{fig:linux}}
    \subfloat[microkernel]{\includegraphics[width=0.435\columnwidth]{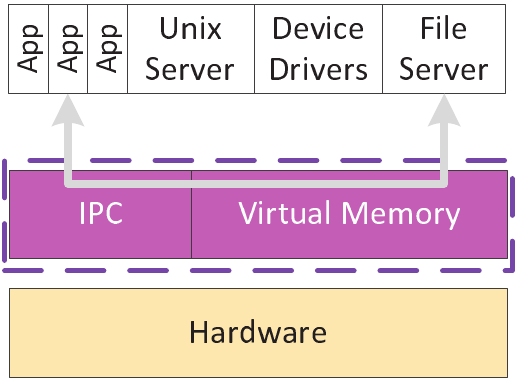}
	\label{fig:micro}}

    \subfloat[exokernel~\cite{Engler1995exokernel}]{\includegraphics[width=0.3\columnwidth]{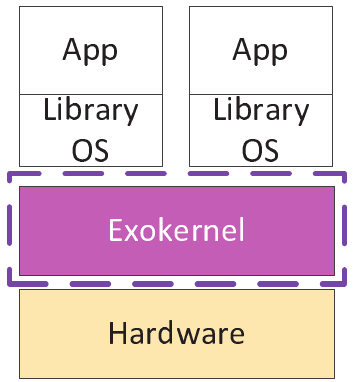}
	\label{fig:exo}}
    \subfloat[unikernel~\cite{madhavapeddy2013unikernels}]{\includegraphics[width=0.33\columnwidth]{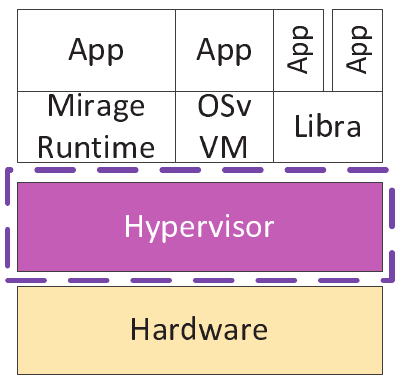}
	\label{fig:uni}}

    \subfloat[multikernel~\cite{Baumann2009Multikernel}]{\includegraphics[width=0.33\columnwidth]{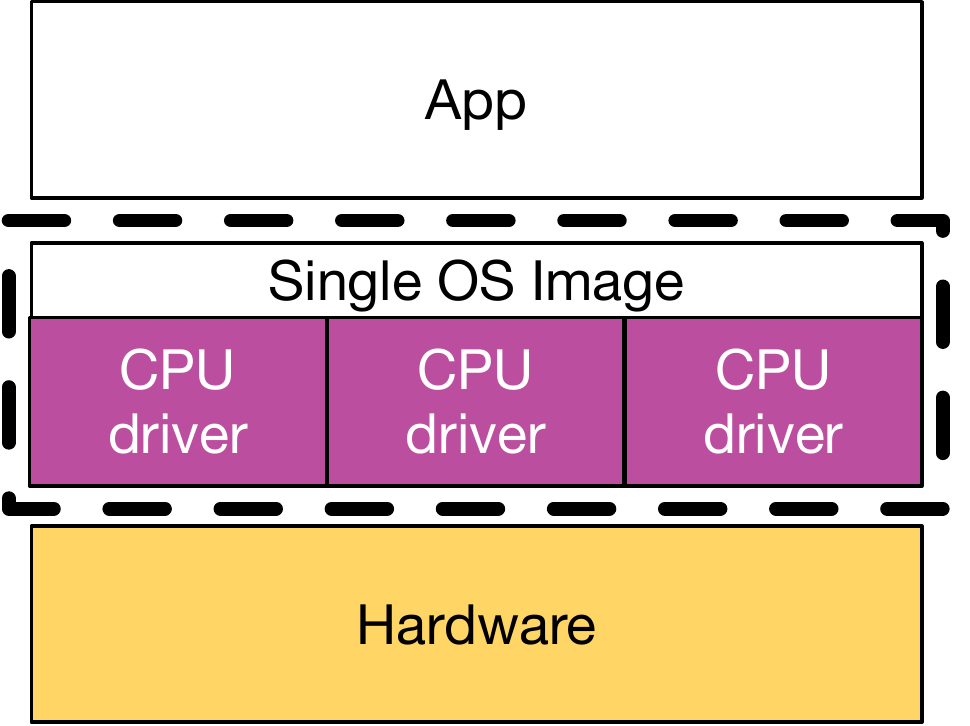}
	\label{fig:nohype}}
    \subfloat[ITFS OS~\cite{rainforest}]{\includegraphics[width=0.33\columnwidth]{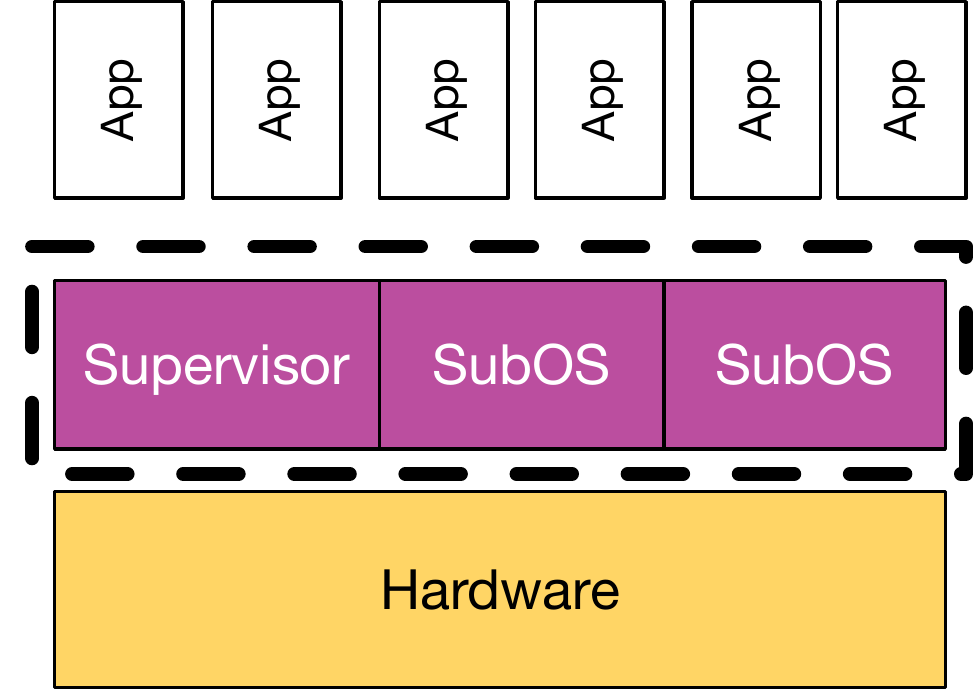}
	\label{fig:rainforest}}

    \subfloat[XOS nokernel]{\includegraphics[width=0.340\columnwidth]{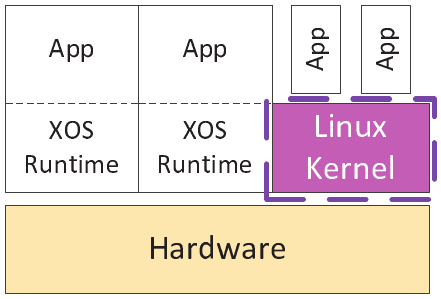}
	\label{fig:xos}}

}
\caption{The difference of the XOS nokernel model from the other OS models.}
\label{XOS_model}
\end{figure}

Modern DCs support increasingly diverse workloads that process
ever-growing amounts of data. To increase resource utilization,
DC servers deploy multiple applications together on one node, but
the interferences among these applications and the OS lower
individual performances and introduce unpredictability.
Most state-of-the-practice and state-of-the-art OSes (including Linux)
were designed for computing environments that lack the support for diversity of resources and workloads 
found in modern systems and DCs, respectively, and they present user-level software
with abstracted interfaces to those hardware resources, whose policies are only optimized for a few specific classes
of applications.

In contrast, DCs need streamline OSes
that can exploit modern multi-core processors,
reduce or remove the overheads of resource allocation and management, and
better support performance isolation.
Although Linux has been widely adopted as a preferred
OS for its excellent usability and programmability, it has limitations
with respect to performance, scalability, and isolation. Such general-purpose,
one-size-fits-all designs simply cannot meet the needs of all applications.
The kernel traditionally controls resource abstraction and allocation, which hides
resource-management details but lengthens application execution
paths. Giving applications control over functionalities usually
reserved for the kernel can streamline systems, improving both
individual performances and system throughput~\cite{Engler95avm:application-level}.

There is a large and growing gap between what DC applications need and
what commodity OSes provide. Efforts to bridge this gap range from
bringing resource management into user
space to bypassing the kernel for I/O
operations, avoiding cache pollution by batching system calls,
isolating performance via user-level cache control,
and factoring OS structures for better scalability on many-core platforms.
Nonetheless, several open issues remain.
Most approaches to constructing kernel subsystems in user space only support
limited capabilities, leaving the traditional kernel responsible for expensive activities
like switching contexts, managing virtual memory, and handling interrupts.
Furthermore,
many instances of such approaches
cannot securely expose hardware resources to user space: applications must load
code into the kernel (which can affect system stability).
Implementations based on virtual machines incur overheads introduced by the hypervisor layer.
Finally, many innovative approaches break current programming paradigms, sacrificing
support for legacy applications.

Ideally, DC applications should be able to
finely control
their resources, customize kernel policies for their own benefit, avoid interference
from the kernel or other application processes, and scale well with the number of
available cores.

From 2011 to 2016, in collaboration with Huawei, we have a large project to investigate different aspects of DC computing, ranging from benchmarking~\cite{jia2013characterizing, wang2014bigdatabench}, architecture~\cite{Ma2015pard}, hardware~\cite{hou2013cost}, OS~\cite{rainforest}, and programming~\cite{he2015hadoop}. Specifically, we explores two different OS architectures for DC computing~\cite{zheng2016method, zheng2016methodb, lu2016resource}. This paper~\footnote{The another OS architecture for DC computing is reported in~\cite{rainforest}} dedicates to XOS---an application-defined OS architecture
that follows three main design principles:
\begin{itemize}
\item
Resource management should be separated from the OS kernel, which then merely provides
resource multiplexing and protection.
\item
Applications should be able to define user-space kernel subsystems that
give them direct resource access and customizable resource control.
\item
Physical resources should be partitioned such that applications have
exclusive access to allocated resources.
\end{itemize}

Our OS architecture provides several benefits. Applications have direct
control over hardware resources and direct access to kernel-specific functions;
this streamlines execution paths and avoids both kernel overheads and those of
crossing between user and kernel spaces. Resource-control policies
can be tailored to the classes of applications coexisting on the same OS. And
elastic resource partitioning enables better scalability and performance isolation
by vastly reducing resource contention, both inside and outside the kernel.
We quantify these benefits using microbenchmarks that stress different aspects
of the system plus several important DC workloads.
XOS outperforms state-of-the-art systems like
Linux and Dune/IX~\cite{Belay:2012:Dune, adam2014ix}
by up to 2.3$\times$ and 3.3$\times$, respectively, for a virtual
memory management microbenchmark. Compared to Linux, XOS
speeds up datacenter workloads from
BigDataBench~\cite{wang2014bigdatabench} by up to 70\%.
Furthermore, XOS scales 7$\times$ better on a 24-core machine and performs 12$\times$
better when multiple memory-intensive microbenchmarks are deployed together.
XOS keeps tail latency --- i.e., the time it takes to complete requests
falling into the 99th latency percentile
--- in check while achieving much higher resource utilization.


\section{Background and Related Work}

XOS is inspired by and built on much previous work in OSes
to minimize
the kernel, reduce competition for resources, and specialize functionalities
for the needs of specific workloads. Each of
these design goals addresses some, but not all, of the requirements for DC workloads.
Our application-defined OS model is made possible by hardware support for virtualizing
CPU, memory, and I/O resources.

Hardware-assisted virtualization goes back as far as the IBM System/370,
whose VM/370~\cite{VM370:1981} operating system presented each user (among hundreds or even
thousands) with a separate virtual machine having its own address space and virtual devices.
For modern CPUs, technologies like Intel\textsuperscript{\textregistered} VT-x~\cite{VTx:2005}
break conventional CPU privilege modes into two new modes: \emph{VMX root mode}
and \emph{VMX non-root mode}.

For modern I/O devices, \emph{single root input/output virtualization}
(SR-IOV)~\cite{kutch2011pci} standard allows a network interface (in this case PCI Express, or PCI-e) to be safely
shared among several software entities. The kernel still manages physical device configuration,
but each virtual device can be configured independently from user level.

OS architectures have leveraged virtualization to better meet
design goals spanning the need to support increasing hardware heterogeneity
to provide better scalability, elasticity with respect to resource allocation,
fault tolerance, customizability, and support for legacy applications. Most
approaches
break monolithic kernels into smaller components
to specialize functionality and deliver more efficient performance and fault isolation
by moving many traditionally privileged activities out of the kernel and into
user space.

Early efforts like HURRICANE~\cite{unrau1995hurricane} and
Hive~\cite{chapin1995hive} were designed for scalable
shared-memory (especially NUMA) machines.
They deliver greater
scalability by implementing microkernels (Figure~\ref{fig:micro}) based on hierarchical clustering.

In contrast, exokernel architectures~\cite{Engler1995exokernel} (Figure~\ref{fig:exo}) support
customizable application-level
management of physical resources. A stripped-down kernel securely exports hardware
resources to untrusted library operating systems running within the applications themselves.
Modern variants use this approach to support Java workloads (IBM's Libra~\cite{Ammons2007libra})
or Windows applications (Drawbridge~\cite{Porter2011Drawbridge}).
Like Libra, avoiding
execution on a traditional Linux kernel that supports legacy applications requires
adapting application software.
Unikernels such as Mirage and OSv~\cite{kivity2014osv} (Figure~\ref{fig:uni})
represent exokernel variants
built for cloud environments.

Early exokernels often loaded application extensions for resource
provisioning into the kernel, which poses stability problems.
Implementations relying on virtual machine monitors can unacceptably lengthen
application execution paths.
XOS avoids such overheads by leveraging virtualization
hardware to securely control physical resources and only uploading streamlined, specialized
kernel subsystems to user space.

The advent of increasingly high core-count machines and the anticipation of future
exascale architectures have brought a new emphasis on problems of scalability.
Corey~\cite{Boyd-Wickizer2008corey}
implements OS abstractions to allow
applications to control inter-core resource sharing, which can significantly improve
performance (and thus scalability) on multicore machines.
Whereas domains in earlier exoscale/library-OS solutions often shared an OS image,
replicated-kernel approaches like Hive's are increasingly being employed.
Multikernel~\cite{Baumann2009Multikernel} employ no virtualization layer, running
directly on bare hardware, without a virtualization layer. The  OS (Figure~\ref{fig:nohype})
is composed of different CPU driver, each running on a single core.
The CPU drivers cooperate to maintain a consistent OS view.

Our another DC OS architecture~\cite{rainforest} introduces the "isolated first, then share" (IFTS) OS model in which hardware resources are split among disparate OS instances. The IFTS OS model decomposes the OS into supervisor and several subOSes, and avoids shared kernel states between application, which in turn reduces performance loss caused by contention.

Several systems streamline the OS by moving traditional functionalities into user
space. For instance, solutions such as
mTCP~\cite{Jeong2014mtcp}, Chronos~\cite{Kapoor2012chronos}, and MegaPipe~\cite{Han2012Megapipe}
build user-space network stacks to lower or avoid
protocol-processing overheads and to exploit common communication patterns.
NoHype~\cite{Keller2010NoHype} eliminates the hypervisor to reduce OS noise.
XOS shares the philosophy of minimizing the kernel to realize performance
much closer to that of bare metal.

Dune~\cite{Belay:2012:Dune}/IX~\cite{adam2014ix} and
Arrakis~\cite{peter2014Arrakis} employ approaches very similar to XOS.
Dune uses nested paging to support user-level control over virtual memory.
IX (Dune's successor) and Arrakis use hardware virtualization to separate
resource management and scheduling functions from network processing.
IX uses a full Linux kernel as the control plane (as in Software Defined
Networking), leveraging memory-mapped I/O to implement pass-through access
to the NIC and Intel's VT-x to provide three-way isolation between
the control plane, network stack, and application.
Arrakis uses Barrelfish as the control plane and exploits IOMMU and
SR-IOV to provide direct I/O access.
XOS allows applications to directly access not just I/O subsystems
but also hardware features such as memory management and
exception handling.

\section{The XOS Model}

The growing gap between OS capabilities
and DC workload requirements necessitates that we rethink
the design of OSes for modern DCs.
We propose an \emph{application-defined OS} model. This OS model is guided by three principles:
1) separation of resource management from the kernel;
2) application-defined kernel subsystems; and
3) elastic resource partitioning.

\subsection{Separating Resource Management}

We contend that layered kernel abstractions are
the main causes of both resource contention and application interference.
Achieving near bare-metal performance thus requires that we
remove the kernel from the critical path of an application's execution.

In an application-defined OS (Figure~\ref{fig:xos}), the traditional role of OS kernel is split.
Applications take over OS kernel duties with respect to resource configuration,
provisioning, and scheduling. This allows most kernel subsystems to be
constructed in user space. The kernel retains the responsibility
for resource allocation, multiplexing, and protection, but
it no long mediates every application operation.
Reducing kernel involvement in application execution has several advantages:
first, applications need not trap into kernel space. In current general-purpose
OSes like Linux, applications must access resources through the kernel,
lengthening their execution paths. The kernel may also interrupt
application execution: for instance, a system call invocation typically
raises synchronous exception, which forces
two transitions between user and kernel modes.
Moreover, it flushes the processor pipeline twice
and pollutes critical processor structures, such as the TLB, branch
prediction tables, prefetch buffers,
and private caches. When a system call competes for shared kernel
structures, it may also stall the
execution of other processes using those structures.

One challenge in separating resource management from the kernel
is how to securely expose hardware to
user space. Our model  leverages modern virtualization-support
hardware. Policies governing the handling of privilege
levels, address translation, and exception triggers can enforce
security when applications directly interact with
hardware. They make it possible for an application-defined OS to give
applications the ability to access all privileged instructions, interrupts, exceptions,
and cross-kernel calls and to have direct control over
physical resources, including physical memory, CPU cores, I/O devices,
and processor structures.

\subsection{Application-Defined Kernel Subsystems}

Our OS model is designed to separate all resource
management from the kernel, including CPU, memory, I/O, exceptions.
Applications are allowed to customize their own kernel subsystems, choosing the
types of hardware resources to expose in user space.
Furthermore, applications running on the same node may implement
different policies for a given kernel subsystem.
Kernel services not built into user space are
requested from the kernel just as in normal system calls.

Application-defined kernel subsystems in user space are a major feature of XOS.
Applications know what resources they need and
how they want to use them. Tradeoffs in traditional OS design
are often made to address common application needs, but this
leads to poor performance for applications without the such ``common''
needs or behaviors. For example, data analysis workloads with
fixed memory requirements will
not always benefit from demand-paging in the virtual memory subsystem, and
network applications often need
different network-stack optimizations for better throughput and latency.

In this model, a \emph{cell} is an XOS process that is granted exclusive
resource ownership and that runs in VMX non-root mode.
Cells can bypass the kernel to have fine-grained direct control over physical
resources. Unlike other user-space I/O approaches that construct I/O stacks in
user space, XOS allows the construction of any kernel subsystem within each cell.
Such subsystems can include paging, physical memory
management, I/O, and interrupt handling, and the application-defined OS model
allows diverse policies for each subsystem. For example, XOS allows applications
to request physical pages in specific colors or banks in order to
reduce cache conflicts and enable better performance isolation.
Our XOS model implements pass-through functionality for PCI-e devices with
the help of SR-IOV,
To avoid device exhaustion, message-based I/O system calls within XOS runtime service I/O requests.

\subsection{Elastic Resource Partitioning}

In the XOS model,  each application has exclusive ownership of the resources allocated
to it, including kernel structures and physical resources. Figure~\ref{fig:xos} shows how
elastic partitioning works. Each cell has unrestricted access to its
resources, including CPU cores, physical memory, block device, NICs, and privileged features.
It controls physical resources by scheduling them in whatever way it chooses. A cell with
exclusive resources runs with little kernel involvement, since it
shares no state with other cells. The kernel only provides
multiplexing, protection, and legacy support.

One obvious drawback to such elastic partitioning is that it prevents
resource sharing, which may reduce utilization.
To overcome this limitation, we build reserved resource pools both in the kernel and
in the XOS runtime.
These pools can grow or shrink as needed.
The kernel's resource pools are per-CPU structures, which ensures
that cells do not contend for resources.
If a cell exhausts its allocated resources, it can invoke XOS runtime routines to request
more resources from the kernel. By carefully accounting for
the resources allocated to each cell,
the kernel tracks resource consumption for each software component.
It can thus guarantee QoS
for each cell through performance isolation and judicious resource allocation.
For instance, the kernel could choose
to devote a fraction of the memory and I/O devices to
a resource pool serving a critical cell.
The need for elastic partitioning grows when co-existing applications
compete for resources,
which is common in real-world deployment. Without effective mechanisms
to provide performance guarantees,
co-existing applications may perform unpredictably, especially
if they are kernel-intensive.

\begin{figure}[tb]
\centering
\includegraphics[width=0.95\columnwidth]{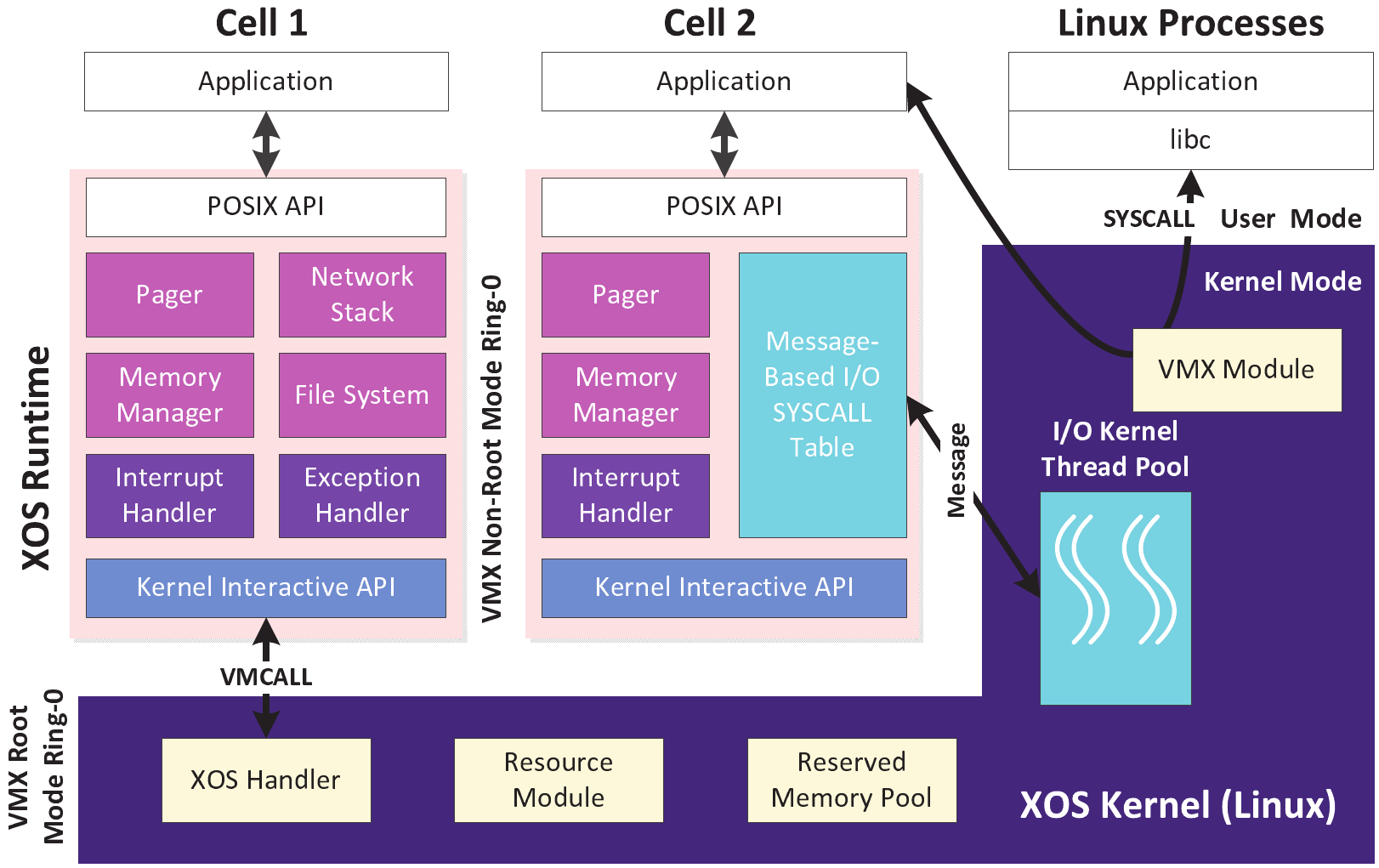}
\caption{The XOS architecture}
\label{XOS_architecture}
\end{figure}

\section{XOS Implementation}

Figure~\ref{XOS_architecture} shows the XOS architecture, which
consists of a number of dynamically loadable kernel modules plus the
XOS runtimes.
The XOS kernel has five more functionalities than an ordinary kernel:
initiating and configuring the
VMX hardware (the \emph{VMX} module); allocating and accounting resources for cells (the \emph{resource}
module); handling XOS runtime requests and violations (the \emph{XOS handler}); handling
message-based I/O system calls (I/O kernel thread pool); and providing a physical memory allocator to
reserve a physical memory pool for XOS processes.

The XOS kernel runs in VMX root mode ring 0,
and the cells run in VMX non-root mode ring 0.
XOS leverages VT-x to enable privileged hardware features
to be securely exposed in VMX non-root mode,
which allows the cells to directly manage hardware resources
without trapping into the kernel.
The XOS handler intercepts VM-exits caused by privilege violations
and VMCALLs initiated by the cells --- this is
the main means by which a cell interacts with the
kernel.

XOS cells can coexist with normal Linux processes.
Each cell has an independent virtual address space and a private XOS runtime.
The resources assigned to each cell are exclusive and cannot be accessed by other cells.
We construct traditional kernel services in user space by inlining kernel subsystems
into the XOS runtime.
These kernel services can be customized to meet the needs of different workloads.
In addition to the
kernel services defined in XOS runtime, other services
can be obtained directly from the kernel
via hardware traps (VMCALLs) or messages.
The runtime wraps user-space OS services with the POSIX API,
which provides compatibility with legacy software.

The XOS runtime is a thin, trusted layer,
that is responsible for resource management and kernel interaction
during resource (re)allocation. It is implemented with statically-linked libraries. We offer
two classes of interfaces: one includes explicit interfaces for direct hardware control,
including pre-allocated pages, colored page allocation,
page table entries,
TLB entry invalidations, I/O control-space access, and DMA management. The other includes
POSIX-like interfaces. It invokes inlined kernel functions customized in XOS runtime, while other
system calls can be
served by the kernel. Specially, when there are too few I/O devices dedicated
to each cell, the XOS runtime provides dedicated message-based I/O interfaces,
redirecting I/O system calls
to I/O system services cell via messages. In such a design, I/O services are
deployed in different kernel
threads. Consequently, the processor structures within cells will not be flushed.

\subsection{Booting a New Cell}
In XOS, converting a normal Linux process into a cell process occurs when the Linux process needs
acceleration in certain kernel subsystems. XOS needs two mode switches to make a cell online.
The most important part is to set up a suitable operating environment before and after the booting.

XOS provides a control interface for applications to apply for resources. With the control interface,
applications could specify exclusive resources and
privileged features. Once an application invokes the control interface, the VMX module initiates
VT-x hardware via ioctl() system call, and the application makes the first mode switch into vmx
root mode. After that, the resource module allocates exclusive resources from the resource pool. Then,
the vmx module uploads the original page table into the allocated memory for further user space page table
initialization. The VMX module constructs new Interrupt Descriptor Table (IDT) for the cell.
If the application reserves I/O devices, IOMMU is configured to map
the allocated memory region accordingly. Meanwhile,
XOS handler in the kernel registers new exception handlers
for the cell, as specified by the application. Finally, XOS duplicates
current processor
states
into VMCS host-state area, and sets the processor state and entry point in
VMCS guest-state area.
The VM-execution control fields of VMCS are also set to identify
the privileged features in user space. In vmx root mode, the vmx module triggers VMLAUCH
instruction to enable processes to run in non-root mode as cells.

\subsection{Memory Management}

As many DC applications need large chunks of memory to save their growing data sets, the memory management
subsystem in the kernel can easily become a bottleneck. During memory allocation, the kernel need perform
multiple actions that may have negative performance implications. For instance, it needs to lock
related structures, such as page table entries, before modifying them. It needs to trigger
 TLB shootdowns to flush associated TLB entries.
When multiple processes make memory allocation
 calls, the ensuing contention could noticeably slow down all processes involved.
XOS differs
with others in several ways with respect to the memory management.
Each cell has its own pager and physical memory, handles virtual memory in  XOS runtime rather than in kernel.
The memory manager shares no states with others. XOS kernel merely allocates, deallocates memory resource, and maintains access control list for the resources.

\textbf{Physical memory management.} XOS implements a two-phase physical memory management. The first phase is
to reserve the physical memory in the kernel to launch XOS cells. The second one is user space memory allocator
in XOS runtimes to deliver memory management service.

We argue that applications could benefit from simple memory allocation with large chunks of continuous memory.
Though fine-grained discontinuous memory may improve resource utilization, it increases complexity in the XOS
runtime memory management module. In Linux buddy allocator, the default largest memory chunk that can be allocated is 4MB
Even if we modify the buddy algorithm to allow larger chunk sizes, it will be divided into
fragmentations after OS boots. Consequently, we modify Linux kernel to reserve memory chunks when OS
boots up. The reserved memory is managed by a buddy allocator in XOS resource module. The
maximum chunk allowed is 1024MB in the memory pool. Furthermore, to avoid lock completion when multiple cells
apply for the memory, we build a per-CPU list memory pool. XOS resource manager allocates, reclaims, and
records the states of memory usage in each XOS runtime. XOS runtime maintains another buddy allocator similar as
the one adopted in the XOS resource module, but with a much smaller maximum chunk. The maximum supported
chunk is 64MB, while the minimum chunk is the base page size. XOS runtime uses its buddy allocator and memory
pool to map smaller parts of memory regions into the cell's address space.

\textbf{Virtual memory management.}
XOS runtime handles virtual memory for cells directly rather than in kernel.
For some applications, the memory requirement
cannot be predicted in advance, so the XOS resource module allocates large continuous memory,
and hands it over to user space buddy allocator. XOS runtime then maintains demand paging policies.
Due to performance considerations, we build both pre-paging and demand paging utilities. An application can
choose which one to use on its own.

XOS uploads page table and other process related kernel data into cell's address space,
and back-up the original one in the kernel. We configure the field bits in VMCS guest-state area,
including control register (e.g., CR3), interruptibility state, and etc. When a normal process becomes a cell
by entering VMX non-root mode, hardware will load processor structures from the VMCS region. Consequently,
a cell will inherit the initial address space from Linux. To ensure correctness, the kernel will mlock() the
already mapped page frame in the original page table, preventing them from being swapped out.

\textbf{User-level page fault handler.} Most page faults occur after a process attempts to access addresses
that are not currently mapped to a physical memory location. The hardware raises a page-fault exception and
traps into the exception handler routine. We set bits in VMCS to make page-fault exception not
causing a vmexit in non-root mode. We replace Linux default IDT with our modified one, which will invoke the
user space page fault handler we register inside XOS runtime. The page fault handler will then construct a new
page table entry, with a page frame from user space buddy allocator.

When a cell needs more memory pages than available ones in its user space memory pool, it will request resource from kernel
by triggering a vmexit. The XOS handler in the kernel will synchronize the two page tables,
serve the request and allocate physical memory from the reserved memory pool.

With ability to access its private page table, applications~\cite{riesen2009kitten} with predictable memory
needs can potentially achieve additional performance gains. Functions such as garbage collection can
benefit from manipulating the page table entries directly. Process live migration can benefit from the
user-level page fault handler.

\subsection{Interrupt Management}
In XOS, the kernel hands over I/O devices into user space, including buffer rings, descriptor queues,
and interrupts. As a result, XOS runtime could construct user space I/O device driver to serve I/O operations.
Applications would have direct access to I/O devices, bypassing entire kernel stack. Applications can
also use customized device drivers for further performance improvement. For instance, we could build aggressive
network stacks for small messages processing. PCI-e devices are initiated in Linux kernel. We use PCI-stub
to reserve devices for given cells. Once a device is allocated to a cell, the descriptor rings and
configuration space of that device are all mapped into the cell's address space. When a cell manages the
device, others cannot touch the device. Particularly, when a PCI-e device has multiple virtual functions
(e.g., SR-IOV), XOS passes through a single queue to each cell.

The physical interrupts derived from PCI-e devices raise challenges to the performance of user space
device manager. When XOS runtime maintains a user space device driver, device interrupts are handled
directly in user space. 
Because handling interrupts in the kernel will change the context of the process
structure, and needs to redirect interrupts to specific cores.
When allocating PCI-e device to the
user space, we deliver interrupts to the CPUs on which the cell runs. We replace the default Linux IDT
with XOS cell's IDT, which is registered in XOS runtime. For interrupts permitted in non-root mode,
the interrupt handler found by IDTR will handle the interrupts directly in user space. For interrupts
not permitted in VMX non-root mode, the hardware will trap into the kernel, and trigger the explicit
kernel interrupt handler. We configure explicitly in XOS kernel which interrupts do not cause a vmexit.
After an interrupt is processed, the interrupt handler in XOS runtime is completed with a write to
the x2APIC MSR. If we set the according MSR bitmap in VMCS, signaling interrupt completion will
not cause a vmexit.

\subsection{Message-Based I/O System Calls}

When there are not enough physical devices or virtual device queues for cells that need dedicated devices,
some cell may suffer from waiting for devices becoming available. Meanwhile, context switches due to I/O system
calls and external interrupts delivered to computing process many also degrade the performance. To address
this issue, we implement message-based I/O system calls to separate the kernel I/O operations from
the normal execution path of applications.

Figure~\ref{XOS_architecture} presents XOS architecture with message-based system calls. XOS is divided into
several parts, with cells and I/O services running in their respective resources.
The I/O services runs
in different CPUs, and are given specific devices.
I/O services are classified into two class: polling service threads and serving threads.
Polling service threads only
poll I/O requests from cells and dispatch them among serving threads. Serving threads receive requests from message queues,
perform the received I/O system calls, and response to the dedicate cells.
In XOS, we attempt to implement I/O threads in non-root mode to serve the message-based system calls with
user space device drivers. In the current implementation, we create kernel threads to serve I/O services.
Once a normal process becomes a cell, shared memory buffer with each I/O serving thread is established.
We modified the standard libc,
hooked the posix I/O system call with message-based I/O syscall, and conduct multiple pthread-like
fibers to serve. Once an I/O system call is invoked, a fiber gets current cell's context, invokes a asynchronous message-based syscall,
and yield the execution environment to next fiber. The message-based I/O syscall writes request messages in the shared memory buffer,
and waits for return code.
To gain best
performance for each cell, at least one exclusive serving thread per cell is created to response system call requests.
As the number of cores increases from one generation to another, we have found it acceptable to bind kernel
threads on separate CPU cores. Optimizing kernel threads deployment is part of future work.


The main challenge of aforementioned implementation is to synchronize the context of a cell.
To perform asynchronous message-based I/O system calls, an I/O serving thread needs the context, including virtual address space structures
and file descriptors, of the requesting cells.
To do that, the processor related data is backed up in kernel thread's address space
and updated
with every new change. An I/O system call message
is contained in fixed size structure to
avoid cache line evictions. It includes syscall numbers,
parameters, status bits, and
data pointed by arguments.

\subsection{Security and Fault Tolerance}



To achieve a comparable level of security, no XOS cell should be able to access the data/funtion of other cells, Linux processes, and the kernel without permission. Note that this does not mean that applications are completely protected, as vulnerabilities in XOS runtime and Linux kernel could still be used to compromise the OS.

To modify or inspect other cell's data, one need to access others' register or memory. Since each cell runs on exclusive cores, the registers can not be accessed. Memory access violations are also mitigated. Since XOS enforces access control list for resources, the only way a cell could access other's physical memory would be to alter the pager in XOS runtime. In XOS kernel, we set up an integrity check derived from the integrity measurement scheme, to collect integrity measurements and compare it with those values signed in kernel, with which to verify that a XOS runtime is running as expected or in trusted states, to ensure the necessary chain of trust.

The behaviors of applications are constrained by XOS runtime. Exceptions in user space, such as div/0,
single step, NMI, invalid opcode, and page fault, are caught and solved by XOS runtime. An XOS cell is considered
as an "outsider" process by the kernel. When a cell crashes, it will be automatically replaced without any rebooting.


\begin{table*}[htb]
\caption{System Calls and Privileged Features (Cycles)}
\centering
{
\sffamily
\small
\begin{tabular}{|c r r r r r r r r r r|}
  \hline
  & null syscall & rdtsc & rdtscp & rdpmc & read cr8 & write cr8 & lgdt & sgdt & lidt & sidt\\
  \hline
  Linux & 174 & 4167 & 4452 & 226 & N/A & N/A & N/A & N/A & N/A & N/A\\ 
  XOS & 42 & 65 & 101 & 134 & 55 & 46 & 213 & 173 & 233 & 158\\ \hline
\end{tabular}
}
\label{bare_metal_evaluation}
\end{table*}

\section{Evaluation}


\subsection{Evaluation Methodology}
XOS currently runs on x86\_64-based multiprocessors.
XOS is implemented in Ubuntu with Linux kernel version 3.13.2.
We choose Linux with the same kernel as the baseline and Dune (IX) as another comparison point of XOS.
Most of the state-of-the-art proposals, like Arriaks, OSV, and Barrelfish, are initial prototypes or built
from bottom up. They does not support current Linux software stacks. Thus
only Dune is chosen in our study because it's built on Linux, as the adversary.

We deploy XOS on a node with two six-core 2.4GHZ Intel XEON E5645 processors,
32GB memory, 1TB disk,
and one Intel Gb ethernet NIC. Each core has two hardware threads and a private L2 cache.
Each processor chip has 12MB L3 cache shared by six on-chip cores and supports VT-x.

\begin{table}[!htb]
\caption{Average Cycles for XOS Operations}
\centering
{
\sffamily
\small
\begin{tabular}{|l r|}
  \hline
  XOS Operations & Cycles\\
  \hline
  Launch a cell & 198846 \\
  Interact with Kernel & 3090 \\ \hline
\end{tabular}
}
\label{xos_operation}
\end{table}

\begin{figure}[!hbt]
{
\centering
    \subfloat[sbrk()]{\includegraphics[width=0.85\columnwidth]{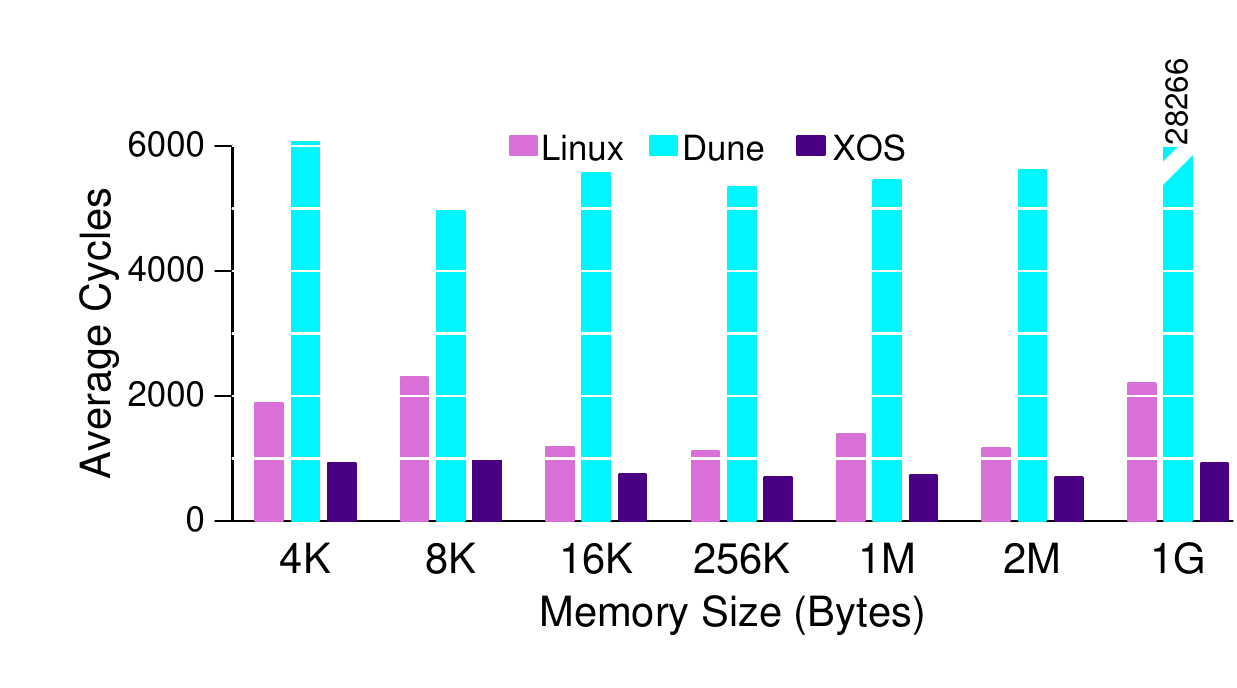}
	\label{fig:sbrk}}

    \subfloat[mmap()]{\includegraphics[width=0.85\columnwidth]{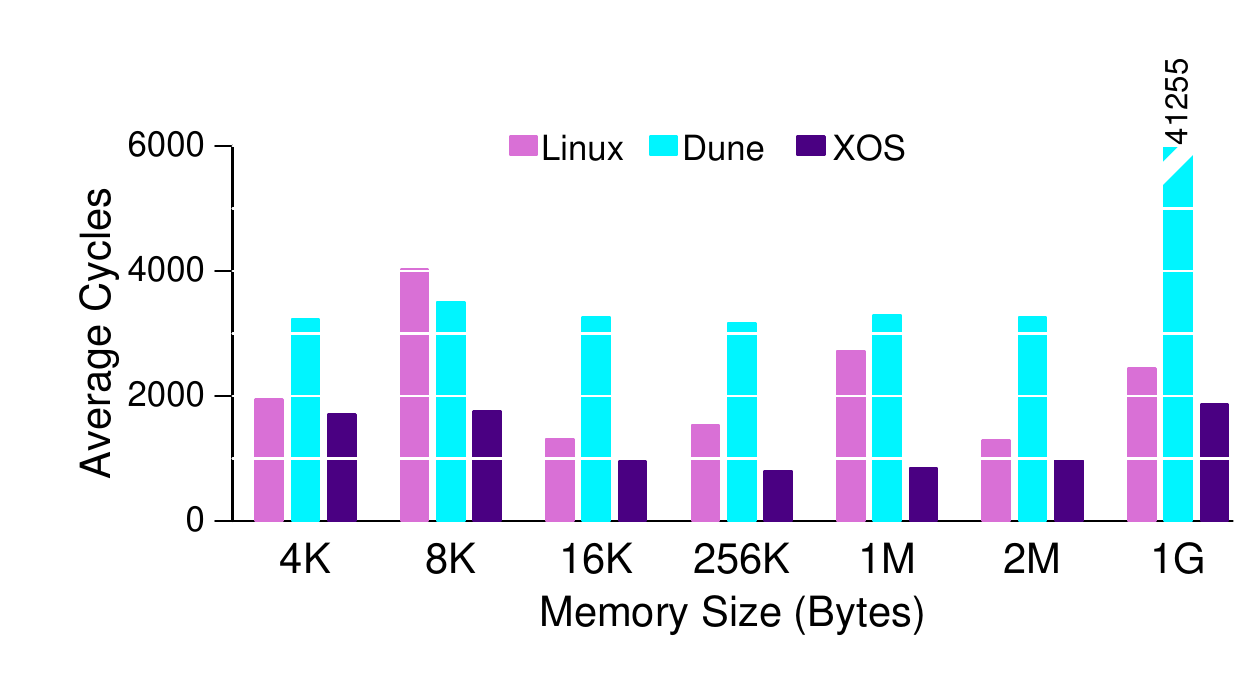}
	\label{fig:mmap}}

    \subfloat[malloc()/free()]{\includegraphics[width=0.85\columnwidth]{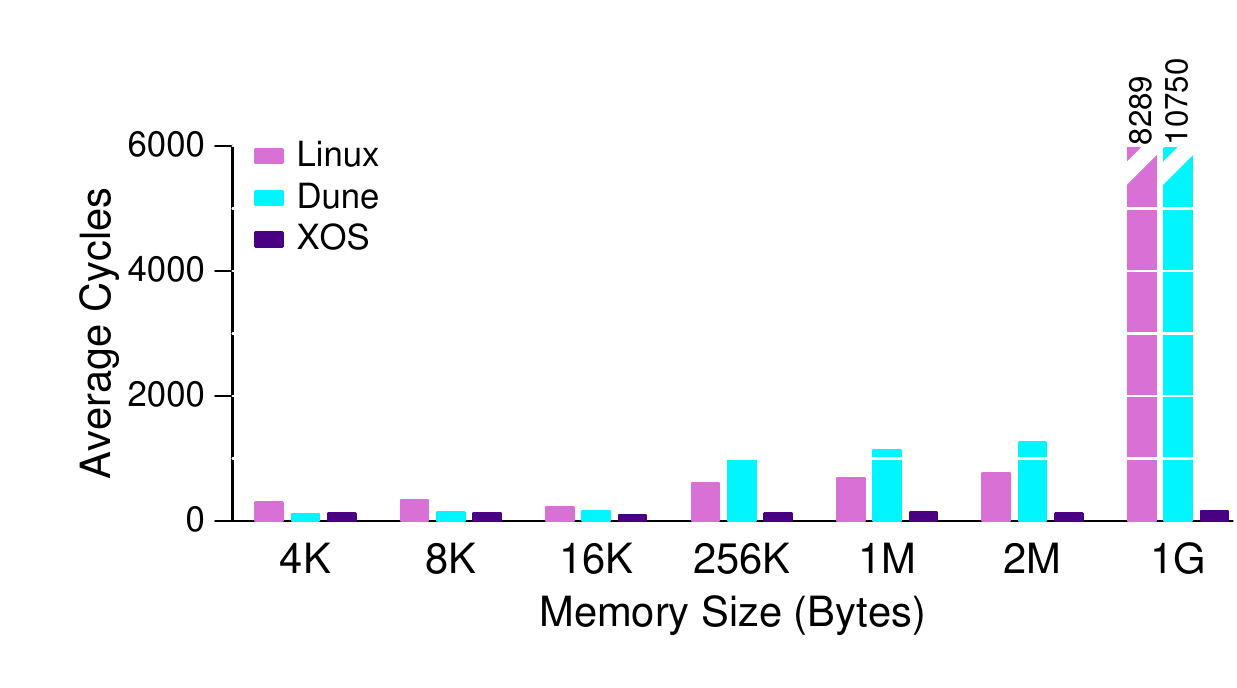}
	\label{fig:malloc_free}}

    \subfloat[malloc()]{\includegraphics[width=0.85\columnwidth]{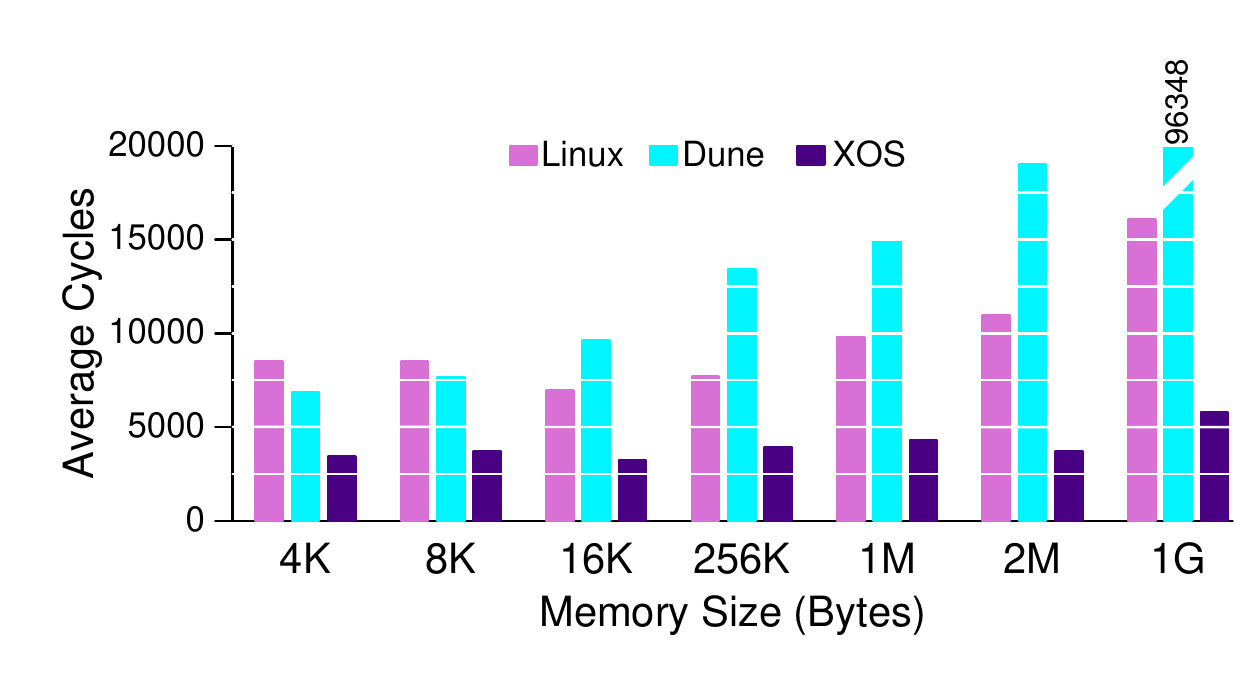}
	\label{fig:malloc}}
}
\caption{Microbenchmark Performance}
\label{micro_perf_on_diff_os}
\end{figure}

\begin{table}[!htb]
\caption{Average Cycles for Memory Access}
\centering
{
\sffamily
\small
\begin{tabular}{|c r r|}
  \hline
   & read & write \\
  \hline
  Linux & 305.5 & 336.0 \\
  Dune & 202.5 & 291.2 \\
  XOS & 1418.0 & 332.0 \\ \hline
\end{tabular}
}
\label{avg_read_write}
\end{table}

We use some well-understood microbenchmarks and a few publicly available application
benchmarks.
All of them are tested on the baseline Linux and XOS. However,
due to Dune not being able to provide robust backward compatibility, only microbenchmarks are able to run on
Dune.
The microbenchmarks also include the Will-It-Scale~\cite{Will_it_scale} benchmark -- 47 typical system calls
for scalability tests,
and the Stress~\cite{stress_benchmark} benchmark -- a performance isolation evaluation tool with
configurable
parameters of CPU, memory, and I/O.
The full-application benchmarks used in our study come from BigDataBench, which includes MPI applications for
data analysis, E-commerce, Search Engine, and SNS.
We run each test ten times and report the average performance figures.

During the evaluation, hyperthreading of the Xeon processor is enabled,
power management features and Intel Turbo Boost are disabled.
The benchmarks are manually pinned to hardware threads, which helps avoid the influence of process
scheduling.
In order to measure precise performance cost, we use rdtsc() to obtain current time stamp of a CPU.
To ensure the sequence of rdtsc() is not optimized by gcc compiler, rdtsc() is defined as \emph{volatile}
to avoid out-of-order compiling.

\subsection{Performance}
To understand the overhead of OS kernel activities, we have conducted experiments that
measure the execution cycles of a few simple system calls.
A null system call is just a system call (e.g., the getpid() system call), that does not invoke other routines.
Others are built to directly execute X86 instructions.
Due to space limitation, we only present the results of rdtsc, rdtscp, rdpmc, read/write cr8, load/store idt and gdt here.
We measure the performance of these system calls on both Linux and XOS runtime,
and categorize X86 instructions into privileged instructions and un-privileged instructions.
Table~\ref{bare_metal_evaluation} shows that a native Linux environment adds an additional 400\% overhead
to the execution time of a null system call.
The overhead in Linux mainly consists of the two mode switching time and other architectural impacts such as
cache flushing.
Comparing to Linux, XOS gains almost 60$\times$ performance in un-privileged instructions, including rdtsc,
rdtscp, and rdpmc.
All of which do not need to trap into kernel when running on XOS.
For XOS, we constructed VMCS regions and implemented simple low-level API in XOS runtime to directly
execute
privileged instructions in user space. As a result,
the user space X86 privileged instructions in XOS shows the similar overhead as the un-privileged ones.


\begin{figure}[tb]
\centering
\includegraphics[width=0.85\columnwidth]{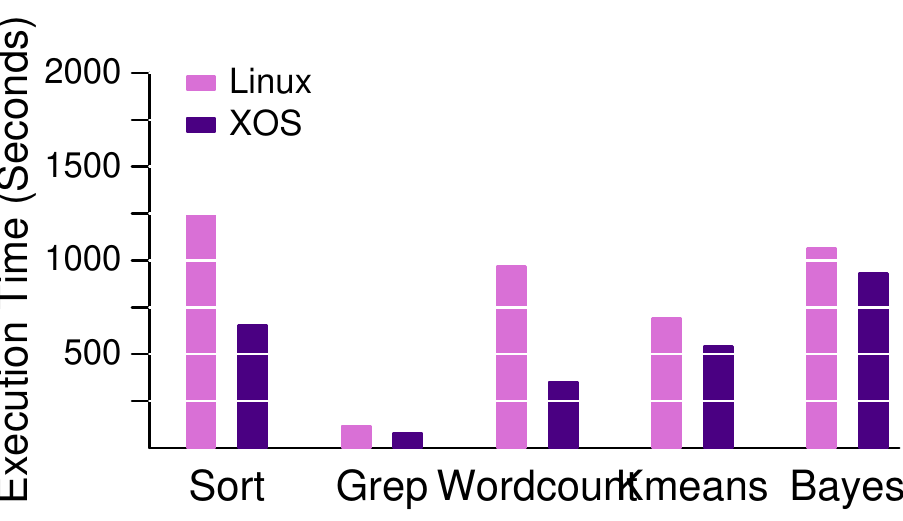}
\caption{BigDataBench Execution Times}
\label{fig:bdbench}
\end{figure}


\begin{figure*}[tb]
{
\centering
    \subfloat[brk()]{\includegraphics[width=0.23\textwidth]{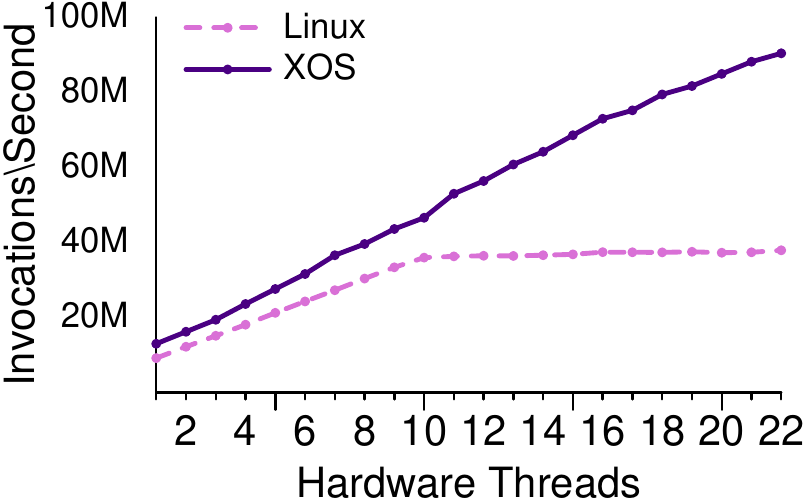}
        \label{fig:brk_scalability}}
     \subfloat[futex()]{\includegraphics[width=0.23\textwidth]{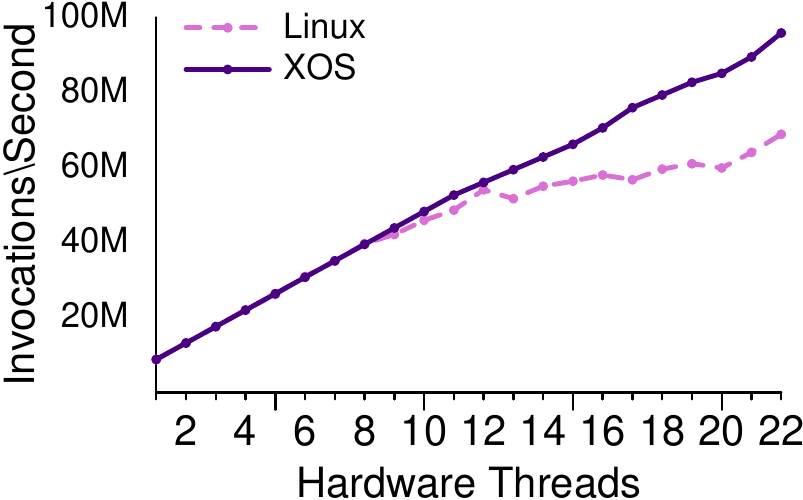}
        \label{fig:futex_scalability}}
    \subfloat[mmap()]{\includegraphics[width=0.23\textwidth]{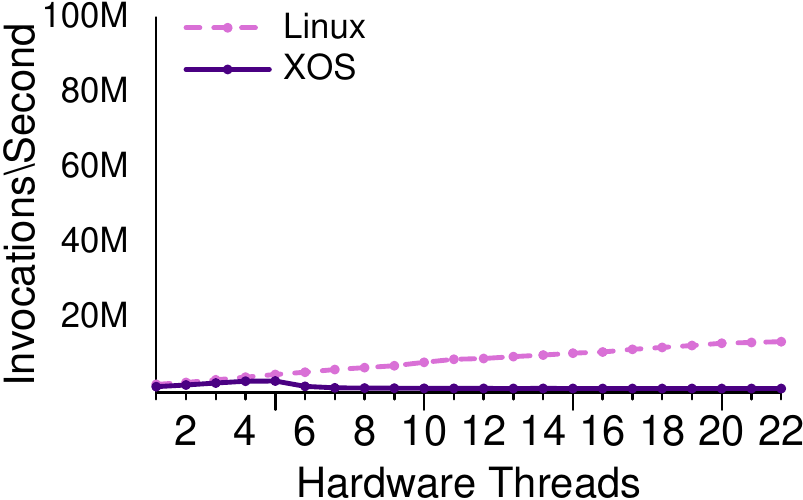}
        \label{fig:mmap_scalability}}
    \subfloat[page fault]{\includegraphics[width=0.23\textwidth]{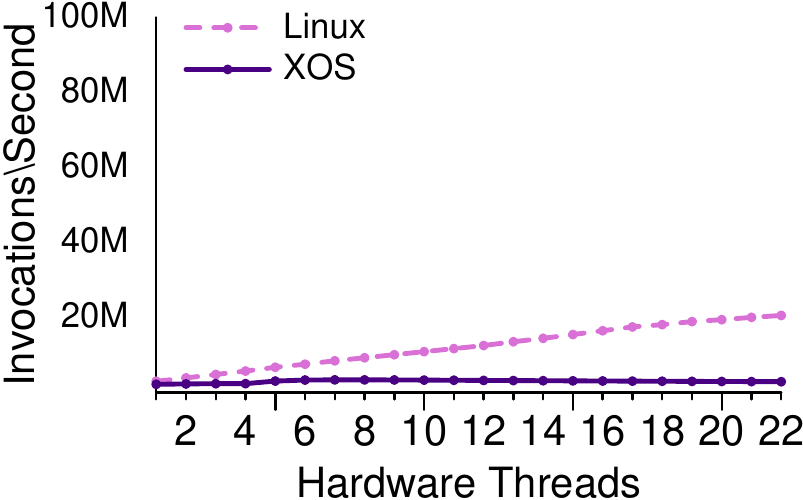}
        \label{fig:page_fault_scalability}}
}
\caption{Scalability Tests}
\label{fig:scalability}
\end{figure*}

\begin{figure}[tb]
\centering
\includegraphics[width=0.85\columnwidth]{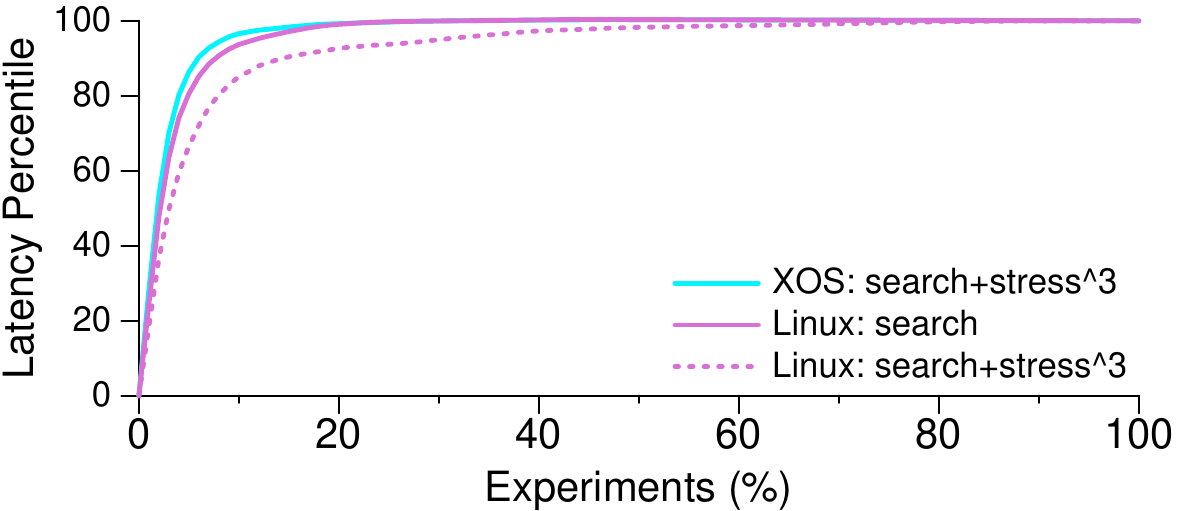}
\caption{Tail latency of Search workload with Interference}
\label{fig:tail}
\end{figure}

To get initial characterization of XOS,
we use a set of microbenchmarks representing the basic operations in big memory workloads,
malloc(), mmap(), brk(), and read/write.
Their data set changes from 4KB (a page frame) to 1GB. Each benchmark allocates or maps a fix-sized memory
region,
and randomly writes each page to ensure the page table is set.

The results are shown in Figure~\ref{micro_perf_on_diff_os}.
We can see that XOS is up to 3$\times$ and 16$\times$ faster than Linux and Dune for malloc() (Figure~\ref{fig:malloc}),
53$\times$ and 68$\times$ faster for malloc()/free() (Figure~\ref{fig:malloc_free}),
3.2$\times$ and 22$\times$ faster for mmap() (Figure~\ref{fig:mmap}),
and 2.4$\times$ and 30$\times$ faster for sbrk() (Figure~\ref{fig:sbrk}).
The main reason is that XOS can provide each process independent memory access ability and user space
resource management,
while Linux and Dune have to compete for the shared data structures in the monolithic kernel.
Moreover, Dune needs to trigger VM-exits to obtain resources from the kernel.
VM-exit is expensive, which causes poor performance.
As the memory size enlarges from 4KB to 1GB, the elapsed time with XOS has no significant change,
while the elapsed time with Linux and Dune increase orders of magnitude.
The main reason is that XOS processes have exclusive memory resource,
and do not need to trap into kernel.
For Linux and Dune, most of the elapsed time is spent on page walks.
A page walk is a kernel procedure to handle a page fault exception, which brings up inevitable overhead
for page fault exception delivery and memory completion in the kernel.
In XOS, page walk overheads are reduced due to user space pagefault handler.
In particular, Linux and Dune experience a significant increase in the execution time of
malloc()/free() (Figure~\ref{fig:malloc_free}).
Malloc()/free() is a benchmark that mallocs fix-sized memory regions, writes to each region, and then frees the allocated memory.
Because XOS runtime hands over the released memory region to the XOS resource pool other than the kernel, and completes all
memory management in user space, which reduces the chance for competitions.
These results prove that XOS can achieve better performance than Linux and Dune.
Table~\ref{avg_read_write} shows that the average time of data read and write in XOS is similar to that in 
Linux,
while Dune has slightly better read performance, but has significant worse write performance.
Dune takes two page walks per page-fault which may cause execution be stalled.

The selected application benchmarks from BigDataBench consist of Sort, Grep, Wordcount, Kmeans, and Bayes.
They are all classic representative workloads in DC.
During the evaluation, the data set for each of them is 10 G.
From Figure~\ref{fig:bdbench}, we can observe that XOS is up to 1.6$\times$ faster than Linux in the
best case.
Compared to the other workloads, Kmeans and Bayes gain less performance improvement, because they are more
CPU-intensive workloads, and do not frequently interact with the OS kernel.
The results prove that XOS can achieve better performance than Linux for common DC workloads.
The better performance is mainly due to the fact that XOS has efficient built-in user space resource
management and
reduces contentions in the kernel.

\subsection{Scalability}

To evaluate the scalability of XOS, we run system call benchmarks from
Will-It-Scale.
Each system call benchmark forks multiple processes that intensively invoke a certain system call.
We test all these benchmarks and find poor scalability of Linux for most of those system calls
implementation.
Some of them have a turning point at about six hardware threads, while others have a turning point at
about 12 hardware threads.
Figure~\ref{fig:scalability} presents the results of brk(), futex(), malloc(), mmap(),
and page faults on XOS and Linux with different number of hardware threads.
The throughput in XOS is better than Linux by up to 7$\times$.
The results show that Linux scales poorly when the core number reaches 15,
while XOS consistently achieves good scalability.
XOS physically partitions resources and bypasses the entire kernel,
thus largely avoids the contentions in the shared kernel structure.
However, please note that a real application will not be so OS-intensive like the system call
benchmarks,
which only reflect the best case scenario for XOS.

\subsection{Performance Isolation}

As XOS targets DC computing,
performance isolation for running coexisted workloads has become a key issue.
The final set of experiments evaluate the performance isolation
provided by XOS architecture.
In this experiment, the system node was setup to run co-existing DC workloads.

The workloads used for these tests were the 
stress benchmarks, and Search from
BigDataBench. Search is a latency-critical workload deployed on three-nodes
cluster. The front-end Tomcat distributes requests from client nodes to back-end Nutch index servers.
With massive tests, we set 150 request/second
in client for tradeoff between throughput and request latency.
We run Nutch servers and stress benchmark in our target OS node,
as Nutch is the bottleneck of Search in our experiment.
The stress benchmark is a multi-thread application where each thread repeatedly allocates
512MB memory and touches a byte per page. In this experiment, we use three-threads stress
benchmark for stability consideration.
We bind each workload on dedicated cores to avoid interference.
The request latencies of all requests are profiled and presented by the cumulative distributions (CDFs)
in Figure~\ref{fig:tail}.
Each request latency is normalized to the maximum latency of all experiments.
The results show that tail latency in XOS outperforms the one in Linux.
Particularly, the 99th latency percentile in XOS is 3$\times$ better than Linux.
In addition, the number of outliers (length of the tails) is generally much smaller for XOS.


\section{Conclusion}
This paper explores the  OS architecture for DC servers. We propose an application-defined OS model guided by three design principles: separation of resource management from the kernel;
application-defined kernel subsystems; and
 elastic partitioning of the OS kernel and physical resources. We built  a Linux-based prototype
to adhere the design model. Our experiments demonstrated XOS's advantages
over Linux and Dune in
performance, scalability, and performance isolation, while still providing full support for legacy code.
We believe that the application-defined OS design  is a promising trend for the increasingly rich variety of DC workloads.





\section*{Acknowledgment}

This work is supported by the National Key Research and Development Plan of China (Grant No. 2016YFB1000600 and 2016YFB1000601). 
The authors are grateful to anonymous reviewers for their insightful feedback.

\bibliographystyle{IEEEtran}
\bibliography{references}


\end{document}